# Convection and Motion in 2-d Embankments Under Cyclic Boundary Conditions

**P. Evesque**
Lab. MSSMat, UMR 8579 CNRS, Ecole Centrale Paris
92295 CHATENAY-MALABRY, France, e-mail: evesque@mssmat.ecp.fr

**Abstract:**
*The motion of grains in a 2d embankment under periodic horizontal forcing is studied theoretically using Coulomb-type modelling. Periodic conditions are used to determined the inclination of the free surface. It is shown that no periodic solution can be found in some domain of the bulk- and wall- friction parameters {$\varphi$, $\varphi_w$} larger than 30°. When a stable periodic solution exists, we show that the finite amplitude of motion leads to generate a flow localised at the free surface, near the bulldozer wall and in the yield band; this may enforce a bulk convection too. At last, we argue why a bulk convection is generated when the periodic solution is not stable. Results are compared to experimental data.*
**Pacs # : 5.40.-a ; 5.45.-a ; 45.70.-n ; 62.20.-x ; 83.50.V**

Let us consider an embankment which is retained by a vertical wall (or a bulldozer) which can move horizontally. Be P the horizontal component of the force exerted by this wall. The embankment is in static equilibrium if the force exerted by the moving wall is larger than a given force $P_{min}$ and smaller than a maximum one $P_{max}$. But as soon as $P<P_{min}$ (or $P>P_{max}$), static equilibrium is no more ensured and the pile moves down (or up) and deforms. One observes in this case that the top part of the medium which is located near the bulldozer slides down (or up) on the bottom part. In general the static zone and the moving one, are well delimited, the last one on top of the other, so that one can define a sliding surface at the interface. One observes also that the inclination of the sliding zone depends on the kind of motion (up or down) [1] and on the inclination of the free surface of the medium [2].

Also, if one starts from an embankment with a horizontal free surface, and enforces the bulldozer to perform a periodic horizontal motion from this state, one observes the free surface which inclines spontaneously whatever the smallness of the motion amplitude. After a while a stationary shape is achieved, and a small convection is observable in the neighbourhood on the bulldozer; one finds also that the convection speed depends on the motion amplitude and frequency. This may surprise; but it is indeed what has been reported already and partially understood, [3].

Teaching soil mechanics leads often to decompose the domain into two distinct fields: (i) the case of small deformation is the domain of foundations for which even quite tiny deformation of soil is important since it can generate major disturbance in the building above it; the case of small deformation is the topics of [4]; (ii) large deformation of soil occurs in case of rupture and flows, for which the earth structure itself becomes in danger. In the second case, one observes often localised deformation with part of the system seeming sliding on the other one. This remarks led Coulomb to describe this domain of the mechanics of soil using an analysis of limit of stability applying a variational method to find the best sliding surface [1]. Since this pioneering work, this technique has been much improved [5].

In the previous experiment of cyclic bulldozing, one gets large deformation and flow by imposing periodic forcing of small amplitude. A question arises then: Is it the domain of large or small deformation? This paper is an attempt to answer this question by describing more precisely the flow observed during the small periodic forcing using a limit analysis and to predict some quantitative effect. The comparison with experimental data shall lead to a better definition of the rheology of granular matter and shall allow concluding if the method is well suited.

The first part of the paper recalls the basic concepts. Section 2 applies the method to understand why the free surface of the embankment progressively inclines and calculate the final inclination. The third part calculates the flow within the above approximations and Section 4 compares the result with experiments. Small periodic forcing generates macroscopic flows in liquid, the process is called "acoustic streaming" in this case. Section 5 discusses some analogies and differences between the two problems of convection (in granular matter and in fluid).

## 1. Some basis of soil mechanics in the domain of large deformation.

Mechanics of granular matter is governed essentially by solid friction and cohesion: Take a container filled with non cohesive sand and inclines it; in short, the sand remains immovable till a maximum slope is reached, then it flows. This exemplifies the effect of solid friction since the maximum slope is approximately the friction angle. (In fact, it is more complicated because the mechanics can be perturbed by dilatancy effect and cohesion [6], that can be even time dependent [7]).

In the case of an embankment of non cohesive material such as the one in Fig. 1, the problem is slightly more intricate, since one shall define two solid friction coefficients $\varphi$ and $\varphi_w$ instead of one, the first one for the bulk; the other one describes the interaction of the bulk with the walls. One can observe experimentally that a large deformation of the medium proceeds very often along the sliding of a zone on top of the other one parallel to the sliding surface. So, to describe the way the system deforms or stay at rest, one can perform a stability analysis of the system in the way proposed by Coulomb [1]; in this case one assumes that either the bulk is at rest when the system of force can be at equilibrium, or that the bulk breaks into two parts, one sliding on the other one along the sliding surface, when the system of force is no more sustainable. The sliding surface is the zone of localised deformation which delimits the two parts; it corresponds to a zone where solid friction is fully mobilised when sliding occurs; and the sliding direction is imposed by the set of external forces, since it imposes the direction of the internal friction forces. Hence, when one knows the sliding surface, one can solve the equation of forces to get the limit of stability.

But this sliding surface is not known in general; and it has to be found. This can be done by optimisation using a set of possible surfaces and finding the less (or most) stable one among the set (depending on the direction of motion). One can repeat the process with other surface sets to get the less (most) stable surface; but one gets already correct results with a single well chosen set.

The method is summed up in Figs. 1 & 2, with sliding surfaces which are inclined planes, in the case of a bulldozer pushing or retaining an embankment. It finds the direction of sliding and it shows that static equilibrium can be satisfied in a whole range of boundary conditions, and a whole range of external forces.

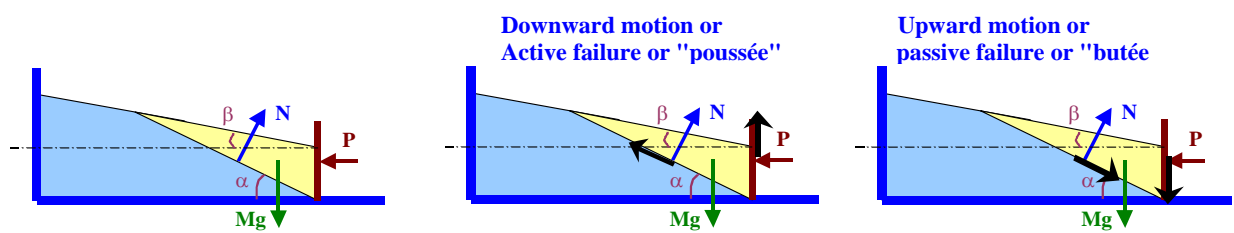

**Figure 1: Notation and process.** $\alpha$= angle between the failure plane and the horizontal. $\beta$= angle of the free surface plane with the horizontal. $\varphi$= bulk friction angle; $\varphi_w$= friction angle between the medium and the moving wall. No cohesion is considered.

P= horizontal component of the force applied by the moving wall to the medium. Mg= weight of the moving part of the medium. $P_t$=P tg($\varphi_w$)=friction force exerted by the moving wall on the medium, at limit of sliding. $N_t$=N tg($\varphi$)=friction force exerted by the non moving part of the medium on the sliding medium, at the limit of sliding. $P_t$ and $N_t$ are directed upward (downward) when the moving part is sliding down (up), i.e. in case of active failure or "poussée" (passive failure or "butée").

**Limit of static equilibrium in the case of downward motion or active failure:**

$N\sin\alpha = P + N\,tg(\varphi)\cos(\alpha)$ and $Mg = P\,tg(\varphi_w) + N\,tg(\varphi)\sin(\alpha) + N\cos(\alpha)$

$\Rightarrow P = N\cos(\alpha)\{tg(\alpha)-tg(\varphi)\}$ and $Mg = P\,tg(\varphi_w) + P[1+tg(\varphi)tg(\alpha)]/\{tg(\alpha)-tg(\varphi)\}$

$\Rightarrow Mg = P\{tg(\varphi_w)+cotg(\alpha-\varphi)\}$ or **$P = Mg/[cotg(\alpha-\varphi)+tg(\varphi_w)]$** =
$P = Mg \sin(\alpha-\varphi)\cos((\varphi_w)/[\cos(\alpha-\varphi+\varphi_w)]$

Solution for $\alpha$ and P in case of active failure: $P = \text{Sup}\{P\} = P_{max} \Rightarrow \partial P/\partial\alpha = 0$

**Limit of static equilibrium in the case of upward motion or passive failure:**

$N\sin\alpha + N\,tg(\varphi)\cos(\alpha) = P$ and $Mg + P\,tg(\varphi_w) + N\,tg(\varphi)\sin(\alpha) = N\cos(\alpha)$

$\Rightarrow P = N\cos(\alpha)\{tg(\alpha)+tg(\varphi)\}$ and $Mg = -P\,tg(\varphi_w) + P[1-tg(\varphi)tg(\alpha)]/\{tg(\alpha)+tg(\varphi)\}$

$\Rightarrow$ **$P = Mg/[cotg(\alpha+\varphi)-tg(\varphi_w)]$** = $Mg \sin(\alpha+\varphi)\cos((\varphi_w)/[\cos(\alpha+\varphi-\varphi_w)]$

Solution for $\alpha$ and P in case of passive failure: $P = \text{Inf}\{P\} = P_{min} \Rightarrow \partial P/\partial\alpha = 0$ or
$0 = \{g[cotg(\alpha+\varphi)-tg(\varphi_w)]\partial M/\partial\alpha + Mg[1+cotg^2(\alpha+\varphi)]/[cotg(\alpha+\varphi)-tg(\varphi_w)]^2$

Firstly, the problem of force equilibrium is solved in Fig.1 assuming a moving zone of mass M, sliding on a surface which is a plane inclined at angle $\alpha$ compared to the horizontal. Here no cohesion is taken into account. It is shown in Fig. 1 caption (i) that the horizontal force P is different for active and passive states, *i.e.* for back and forth motion, (ii) that it depends on the inclination $\alpha$, (iii) that it depends on the mass

M, hence on the shape of the free surface. Fig.2 considers the case of an inclined free surface, for which it describes how M depends on the pile shape.

Following Coulomb procedure, the limit of passive (active) failure is found by finding the minimum (maximum) of P for all possible sliding surfaces and upward (downward) motion. For sake of simplicity the problem is restricted to a set of sliding surfaces . (Also a more complete description of this problem can be found in [2], with a series of applications; and a much more complete discussion on the mathematics underlying the method can be found in [5].) One finds in general that the two optimal failure surfaces, corresponding to active and passive failure, are different. Among the difficulties of the method, it is worth quoting the fact that the sliding direction can be not tangent to the sliding surface [5]. Also, the deformation can be not localised; or the sliding surface can change discontinuously after a while…; the last case is often due to boundary conditions that change continuously, forbidding a constant direction of sliding. These points will not be further discussed here.

To illustrate briefly the method, the case of an embankment with (i) a horizontal free surface, *i.e.* $\beta=0$, (ii) no wall friction, *i.e.* $\varphi_w=0$, is taken now. The set of tested surfaces for the sliding surfaces will be limited to inclined planes of different inclination $\alpha$ and passing through the bottom of the bulldozer wall. So the inclination $\alpha$ being given, and according to captions of Figs.1 & 2, one gets that passive failure is such as $P_{pass}= Mg/[cotg(\alpha+\varphi)-tg(\varphi_w)]$ is minimum and that active failure is such as $P_{act}=Mg/[cotg(\alpha-\varphi)+tg(\varphi_w)]$ is maximum, with $M= \rho h_o^2/[tg(\alpha)-tg(\beta)]$. This leads to:

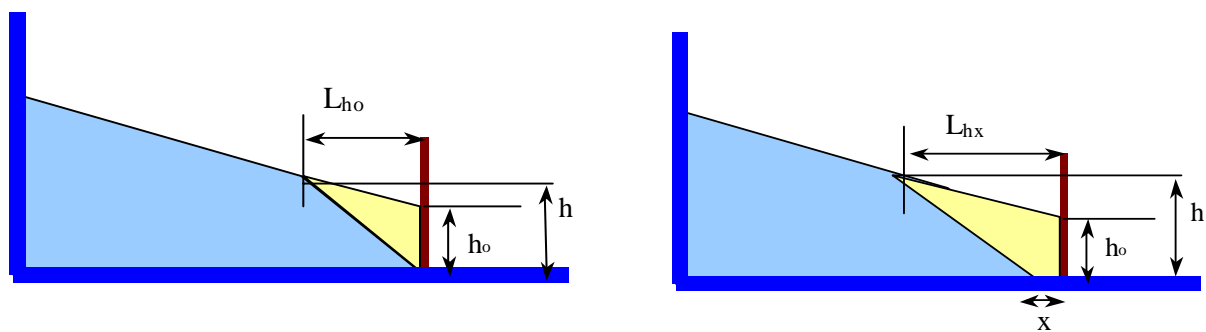

**Figure 2: Mass of the moving zone when the localisation is a plane and the free surface is inclined :**
Be $\alpha$ the localisation angle, be $\beta$ the angle of the free surface slope, be $h_o$ the height of the pile at the moving wall, be $\rho$ the pile density. When the moving zone starts from the bottom of the wall, its mass is given by

$$Mg=\rho g h_o^2/\{2[tg(\alpha)-tg(\beta)]\}$$

since the horizontal extension $L_{ho}$ of the moving zone is given by $L_{ho}=h_o/[tg(\alpha)-tg(\beta)]$ , the surface S of the moving zone is given by $S= L_{ho} L_{ho}[tg(\alpha)-tg(\beta)]/2=h_o^2/\{2[tg(\alpha)-tg(\beta)]\}$.

**Right:** In the same way, if the localisation band starts from a point at x from the bottom of the moving wall, one gets $L_{hx}=L_{ho}+x$, leading to

$$M'g=(\rho g/2)\{(L_{ho} +x)(L_{ho}+x) [tg(\alpha)-tg(\beta)]-x^2/tg(\alpha)\}$$

$$P_{act}= \text{Max}_\alpha \{\rho gh_o^2 \, tg(\alpha-\varphi)/[2tg(\alpha)]\} \quad (1.a)$$

$$P_{pass}= \text{Min}_\alpha \{\rho gh_o^2 \, tg(\alpha+\varphi)/[2tg(\alpha)]\} \quad (1.a)$$

Now it remains to find the two inclinations $\alpha_{min}$ and $\alpha_{max}$ that minimises $P_{pass}$ and maximises $P_{act}$ respectively. Indeed, $\alpha_{min,max}$ can be found by derivation [8], $\partial P/\partial\alpha=0$; this leads to the well-known solutions:

$$\alpha_{act}=\pi/4+\varphi/2 \quad \text{and} \quad P_{act}=[\rho gh_o^2/2][1-\sin(\varphi)]/[1+\sin(\varphi)] \quad (2.a)$$

$$\alpha_{pas}=\pi/4-\varphi/2 \quad \text{and} \quad P_{pass}=[\rho gh_o^2/2][1+\sin(\varphi)]/[1-\sin(\varphi)] \quad (2.b)$$

This is schemed in Fig. 3.a. A question arises then: what does happen to the embankment if one applies cycling bulldozing, *i.e.* back and forth motion? Does the pile shape evolve? An attempt of simulation is sketched in Fig. 3 ; it shows the formation of an inclined free surface. The result is explained in the next section.

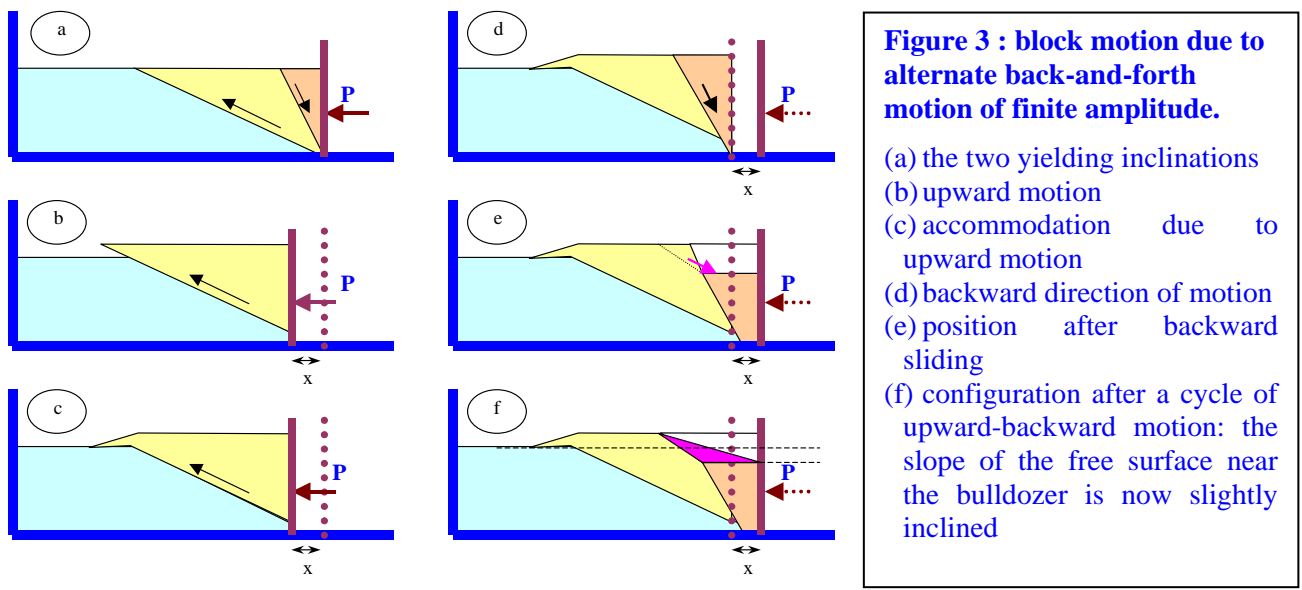

**Figure 3 : block motion due to alternate back-and-forth motion of finite amplitude.**

(a) the two yielding inclinations
(b) upward motion
(c) accommodation due to upward motion
(d) backward direction of motion
(e) position after backward sliding
(f) configuration after a cycle of upward-backward motion: the slope of the free surface near the bulldozer is now slightly inclined

## 2. Formation of a permanent inclined embankment.

One can understand the generation of the inclined slope at the free surface in the following way: consider the embankment with a horizontal free surface, with the bulldozer on the right, as in Fig. 3.

Pushing the bulldozer inward, *i.e.* towards the medium, as in Fig. 3b, forces the right part of the medium to slide on top of the left one with an angle $\alpha=\pi/4-\varphi/2$. As the amplitude of motion is small but finite, say x, the moving medium rises up, say $\delta h=x \, tg(\alpha)$, and the top left part of this zone is deformed to lay on the top surface of the non moving zone, as indicated in Fig. 3c.

When the bulldozer is pulled backward, *i.e.* towards the right as in Fig. 3d, the slope of sliding is larger, *i.e.* $\alpha'=\pi/4+\varphi/2$, leading to a decrease of height $\delta h'=x \, [tg(\alpha)-$

tg($\alpha'$)], so $\delta h'$= -2 x tg($\varphi$), *cf.* Fig. 3e. As $\delta h'<0$, a surface flow is generated on the right top part of the non moving zone of the active motion in order to keep the free surface inclined less than the friction angle, *cf.* Fig. 3f.

So the geometry of the embankment evolves slowly. One can then ask: does a steady shape exist, in the limit of small amplitude of motion? One finds it experimentally, where it looks as in Figure 4. Indeed, in the preceding model, the stationary shape is obtained if it is possible to find an inclination $\beta_o$ of the free surface such as it corresponds to the same value of $\alpha$ in passive and active motions. So the question becomes: Does such an inclination $\beta_o$ of the free surface slope exist, for which the angle $\alpha$ is the same at passive and active failure? According to captions of Figures 1 and 2, this implies that $\alpha$ corresponds to the minimum (with respect to $\gamma$) of $P_1$=Mg/[cotg($\gamma$+$\varphi$)-tg($\varphi_w$)] , and to the maximum (with respect to $\gamma$) of $P_2$=Mg/[cotg($\gamma$-$\varphi$)+tg($\varphi_w$)] , with Mg=($\rho g h_o^2/2$)/[tg($\gamma$)-tg($\beta$)] .

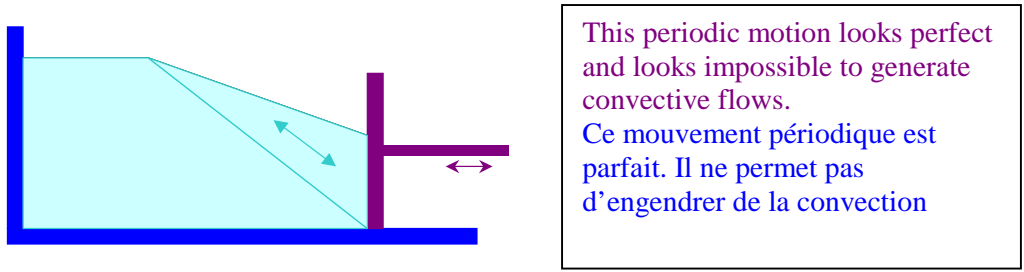

**Figure 4 :  Embankment under cyclic motion of the bull-dozer, in the limit of infinitely small amplitude.** In this case one can find a cyclic solution with an inclined slope of the free surface that ensures the same position and the same inclination of the localisation zones for passive and active state. However, this solution exists only in some range of parameters {$\varphi$, $\varphi_w$} (see text).

So, this implies to find $\alpha$ that is solution of both $\partial P_1/\partial \gamma$=0 and of $\partial P_2/\partial \gamma$=0 at the same time; this writes:

$$\partial\{[tg(\gamma)-tg(\beta)]^{-1}[cotg(\gamma+\varphi)-tg(\varphi_w)]^{-1}\}/\partial\gamma=0 \quad \text{and}$$

$$\partial\{[tg(\gamma)-tg(\beta)]^{-1}[cotg(\gamma-\varphi)+tg(\varphi_w)]^{-1}\}/\partial\gamma=0$$

This leads to the two equations:

$$[1+tg^2(\alpha)]/[tg(\alpha)-tg(\beta)]=[1+cotg^2(\alpha+\varphi)]/[cotg(\alpha+\varphi)-tg(\varphi_w)] \quad (3.a)$$

$$[1+tg^2(\alpha)]/[tg(\alpha)-tg(\beta)]=[1+cotg^2(\alpha-\varphi)]/[cotg(\alpha-\varphi)+tg(\varphi_w)] \quad (3.b)$$

Eq. 3.a and Eq. 3.b have to be satisfied at the same time. Combining Eqs. 3.a and 3.b allows to get an equation without $\beta$ : [1+cotg²($\alpha$+$\varphi$)]/[cotg($\alpha$+$\varphi$)-tg($\varphi_w$)]=[1+cotg²($\alpha-\varphi$)]/[cotg($\alpha$-$\varphi$)+tg($\varphi_w$)] ; this leads to:

$$[1+cotg^2(\alpha+\varphi)][cotg(\alpha-\varphi)+tg(\varphi_w)]=[1+cotg^2(\alpha-\varphi)][cotg(\alpha+\varphi)-tg(\varphi_w)] \quad (4.a)$$

While Eq. 3.a can be written:

$$tg(\beta) = tg(\alpha) + [1+tg^2(\alpha)] [tg(\varphi_w) - cotg(\alpha+\varphi)]/[1+cotg^2(\alpha+\varphi)] \quad (4.b)$$

So knowing $\varphi$ and $\varphi_w$, one can use Eq. 4.a to get $\alpha$, then use Eq. 4.b to get $\beta$; hence the problem is solved. However, some limitation exists since the solutions shall satisfy firstly $\beta<\varphi$, and secondly $\alpha>\beta$. Indeed, first condition ensures that the slope of the free surface is stable, *i.e.* smaller than the maximum angle of repose, and second condition ensures that the mass of the moving block is finite (the mass of the moving block becomes infinite when considering cases with $\alpha<\beta$).

Figs. 5.a and 5.b report the variations of the inclination $\alpha$ and of the slope $\beta$ at the free surface, computed from Eqs. 4, as a function of the bulk friction $\varphi$, and for different values of the friction angle $\varphi_w$ between the wall and the medium. One sees that the slope $\beta_{cyclic}$ is most of the time smaller than the friction angle $\varphi$, (the case $\beta_{cyclic}=\varphi$ is sketched by the dashed blue line in Fig. 5b). However, comparison of Fig 5a and 5b demonstrates that large values of $\varphi$ and $\varphi_w$ are not possible since they lead to consider cases with $\alpha<\beta$. The limit of validity domain is given in Fig. 5 caption.

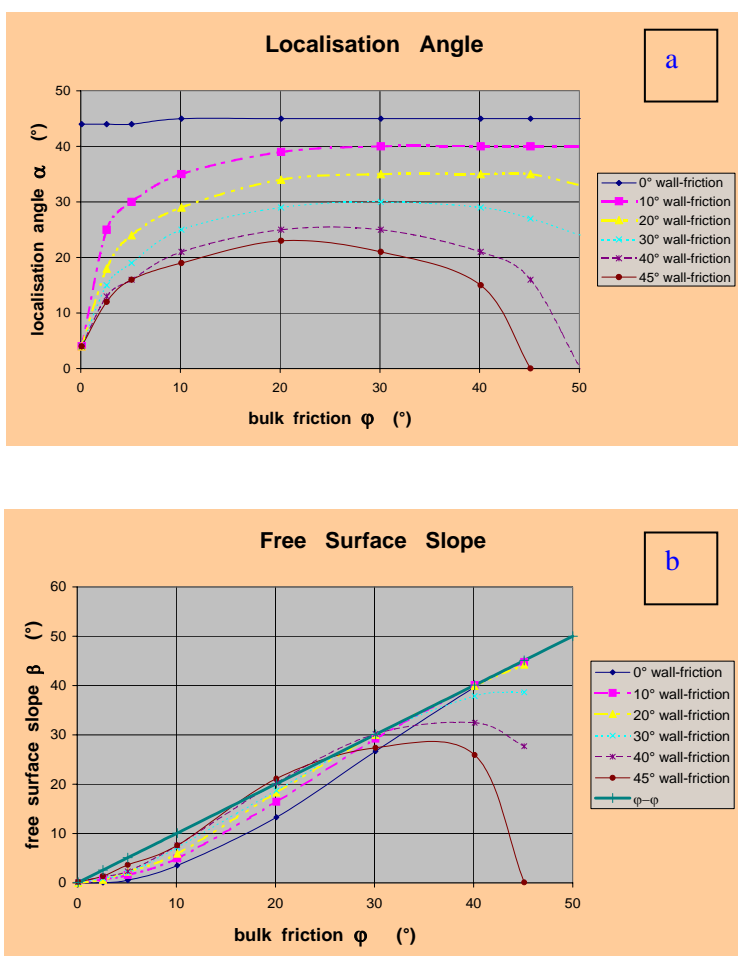

**Figure 5 : Prediction of embankment shape under Cyclic conditions:** Dependence of the inclinations $\alpha$ of the failure zone (a) and of the free surface angle $\beta$ (b) as a function of the bulk friction angle $\varphi$, for different values of the bulk-wall friction $\varphi_w$, i.e. $\varphi_w=0°, 5°, 10°, 15°, 20°, 25°, 30°, 35°, 40°, 45°$.

As $\beta$ is the inclination of the free surface, it shall be such as $\beta<\varphi$. So, the part of the curves that pertains only to the bottom triangle are allowed. This seems requirement to be always satisfied, except for ($\varphi=20°$, $\varphi_w=20°$).

ATTENTION: Solutions such as the slope of the free surface $\beta$ is larger than the inclination of the localisation angle are not possible, because it generates infinite moving mass. This forbids the sets of solution ($\varphi>30°$ and $\varphi_w>20°-30°$) and the sets ($\varphi>40°$ with $\varphi_w>10°$).

These calculations have been confirmed by direct computation of the active and passive forces required to maintain the equilibrium as a function of $\alpha$ and for different values of $\varphi$, $\varphi_w$ and $\beta$. This allows to study the speed of variation of $P_{active/passive}$ with $\alpha$, so the sensitivity of the yielding direction. A typical example is given in Fig. 6, for

φ=30° and φ$_w$=10°. Fig. 6 shows that P$_{passive}$ varies quite fast when the free surface is inclined, which means that the localisation direction is well defined for passive failure.

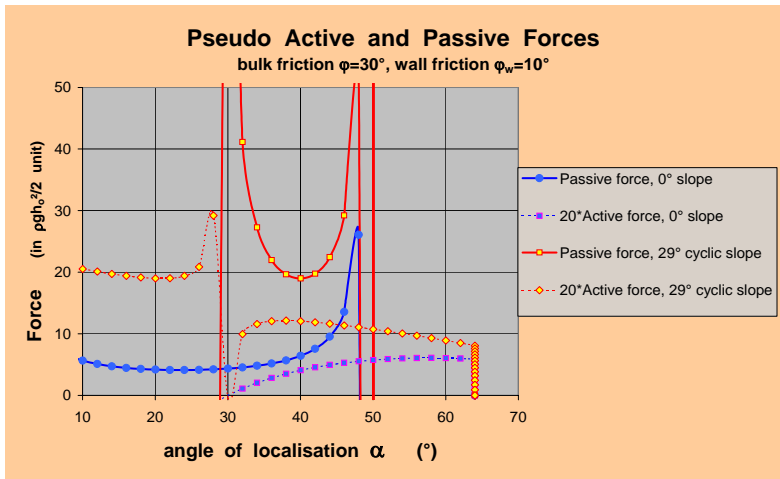

**Figure 6:** Variations of P$_{active}$ and P$_{passive}$ with the angle α of inclination of localisation for a given set of friction angles {φ=30°, φ$_w$=10°}, and 2 different surface inclinations, β=0° and β=β$_{cyclic}$=29°, since one gets β$_{cyclic}$=29° for {φ=30°, φ$_w$=10°}. The minimum of passive failure is much larger when β=β$_{cyclic}$=29° than when β=0°; also Max of active failure and Min of passive failure occur both at α=40° for 29°.

| φ ; φ$_w$ | β=0 Min of P passive | β=0 Max of P active | β$_{cyclic}$ (°) | β$_{cyclic}$ Min of P passive | β$_{cyclic}$ Max of P active |
|---|---|---|---|---|---|
| φ =20°; 0°=φ$_w$ | 2,03 | 0,49 | 13 | 2,79 | 0,61 |
| φ =20°;10°=φ$_w$ | 2,60 | 0,44 | 16 | 4,50 | 0,62 |
| φ =20°; 30°=φ$_w$ | 4,37 | 0,37 | 19 | 16,40 | 0,69 |
| φ =30°; 0°=φ$_w$ | 3,00 | 0,33 | 27 | 7,70 | 0,55 |
| φ =30°; 10°=φ$_w$ | 4,10 | 0,30 | 29 | 19,00 | 0,60 |
| φ =30° ; 20°=φ$_w$ | 5,8 | 0,28 | 29,5 | 79,00 | 0,63 |
| φ =30°; 30°=φ$_w$ | 22,40 | 0,17 | 30 | | |
| φ =40°; 0°=φ$_w$ | 4,61 | 0,22 | 40 | 75,00 | 0,58 |
| φ =40°;10°=φ$_w$ | 6,85 | 0,20 | 40 | | |
| φ =40°; 30°=φ$_w$ | 22,00 | 0,17 | 40 | | |

**Table 1:** This table reports the values of passive and active forces P needed to push (passive) or to retain the wall (active state), when the free surface is horizontal (columns #2 & 3), and when the free surface is inclined at the angle β$_{cyclic}$ which corresponds to stable cyclic conditions (β$_{cyclic}$ is in column #4, passive and active values of P are in columns #5 & 6), for the pairs of friction angles (φ,φ$_w$) for which β$_{cyclic}$ exist. The angles of friction φ and φ$_w$ of the bulk and for the wall-medium interface are given in column #1. The unit of P is ρgh$_o$²/2.
One notes the large increase of the force needed for passive failure when the free surface gets inclined. As noted also in Fig. 5 caption, the cyclic conditions cannot be obtained for: (φ=30°, φ$_w$=30°), (φ=40°, φ$_w$=10°), (φ=40°, φ$_w$=30°)

In contrast, the curve for P$_{active}$ is much flatter; so the direction of active localisation is defined with much less accuracy. Also, Fig. 6 exemplifies that inclination α$_{passive}$ of passive yielding increases when the slope β of the free surface gets more inclined, while inclination α$_{active}$ of active yielding decreases when the slope β of the free surface gets more inclined. This will be important in the next section when studying the causes of convection flow.

An other point which is worth noting on Fig. 6 is the important strengthening of $P_{passive}$ when inclining the free surface slope. This is studied in more details in Table 1, which reports typical values of $P_{active}$ and $P_{passive}$ for different pairs {φ, φ$_w$} and for β=0° and β=β$_{cyclic}$. Indeed one finds a strong "hardening" of $P_{passive}$ due to cyclic forcing, *i.e.* from 4 to 16 for {φ=20°, φ$_w$=30°} or from 6 to 79 for {φ=30°, φ$_w$=20°}. This hardening is not due to the bulk change of mechanical characteristics, **but just by the change of shape** of the embankment. Of course this is due to the change of mass of the zone which moves in front of the bulldozer: this mass increases when β increases, even when one keeps constant the height h$_o$ of the embankment just nearby the bulldozer.

At last, when the cyclic condition leads to β<φ, one can think that the system of back and forth motion is stable and does not generate convection. However the demonstration of this hypothesis requires to investigate carefully the motion of the bulk during cyclic bulldozing , when the cycles have finite amplitude, say x.

## 3. Convection in a permanently inclined embankment due to finite amplitude of forcing.

In caption of Figure 1, the equations have been settled in terms of the mass M of the moving zone; but this mass M varies with the expected direction α of yielding; this has been taken into account in the calculation of the minimum and maximum forces required to let the embankment yield. Doing so, this allows to calculate α, *i.e.* the direction of yielding.

However, the problem becomes more complex as soon as the motion amplitude, x, is finite. As exemplified in Fig. 7, when the bulldozer moves, the bulk deforms; hence the geometry of the embankment varies and the mass M varies too; this changes the condition of localisation; hence this modifies the position of the localisation band and its direction; so it modifies the direction of motion of the grains in the bulk and the size and shape of the moving zone: this means for instance, that some new grains, which were initially pertaining to the immobile (mobile) zone, start moving (or stop moving) after a while, just because they start entering (going out) the localisation zone, while they were out (in) of it just a while before. This effect leads to generate some intermittence and some gradient of motion, which may generate some flow. It is the topic discussed in this section. It will be demonstrated also in the following that the present modelling leads to predict a convection that occurs as soon as periodic bulldozing generates plastic deformation, with a speed which scales linearly with the amplitude x of the motion at first order.

As a matter of fact, Fig. 7 describes the set of problems encountered when one assumes a constant localisation band as soon as the displacement x of the bulldozer gets a finite amplitude. For instance in the case of passive failure, as soon as x is non zero, the bulldozer hits the bottom of the non moving zone, forcing it to deform (problem #2 of Fig. 7) . Also, the medium moves up and the top part of the moving zone shall deform to be supported by the non moving medium (problem #1 of Fig. 7); this modifies the mass M and the inclination $\alpha$ ; so both have to vary with x: $\alpha(x)$, $M(x)$; also, the change of shape modifies the free surface, and its average value $\beta(x)$ is diminished . This is why one expects $\beta(x)$ and $\alpha(x)$ to decrease, and $M(x)$ to increase when x increases, in the case of passive motion.

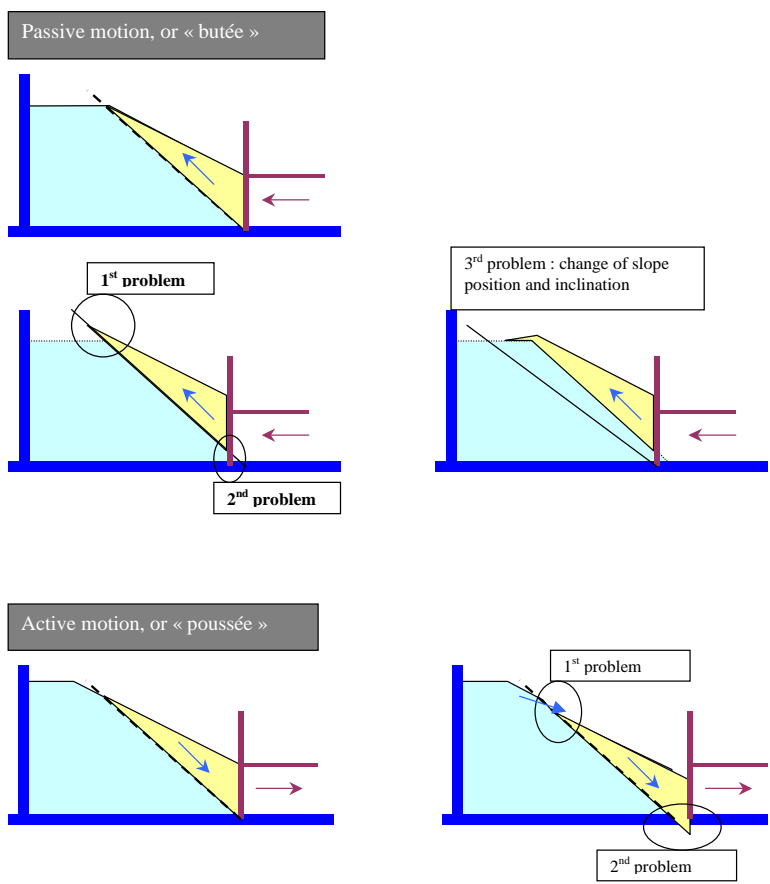

**Figure 7: The sliding motion of the top zone generates different effects during failure:**

**In the case of passive failure** for instance: (i) the top left part of the sliding block is no more supported; this forces the pile geometry to change ; (ii) incompatibility of motion occurs also at the bottom right part, near the wall; (iii) due to (i) the position o the localisation band shall move deeper in the pile, while its mean slope shall decreases because the mean slope of the free surface decreases.

**In the case of passive failure :** (i) As the inclination of the localisation is larger than the friction angle, the sliding generates also a surface flow; (ii) there is also an incompatibility of motion in the bottom right corner; (iii) The localisation surface shall moves out towards right.

Similar problems arise in the case of active failure, as reported in Fig. 7, with a problem of mis-adaptation of the deformation at the bottom of the embankment. Furthermore as $\alpha>\varphi$ in this case, one expects that the new free surface which is generated at the upper limit of the moving zone becomes unstable due to the back motion of this mobile zone, because it generates a free surface which is inclined at $\alpha$ (and $\alpha>\varphi$). This generates some surface flow which modifies the surface shape. In mean, one expects the mean free surface to get more inclined. This lets predict $\beta(x)$ to increase, and $\alpha(x)$ to decrease as x goes backward, in the case of passive motion.

It is possible to propose some way of accommodation of the deformation at the bottom of the embankment near the bulldozer. Two possible processes are sketched in Fig. 8, one for passive failure, the other one for active failure. As shown in this figure, they are both based on a slight variation of the position of the failure zone. This variation shall be quite small otherwise it would modify importantly the mass of the moving zone, leading to (i) the need of too much extra force in case of passive failure, or to (ii) a too much important decrease of the needed retaining force. Anyhow, Fig. 8 shows that special flow can occur at this corner; this let predict a convection flow different from the mean in this zone. This is indeed what has been reported in [3b].

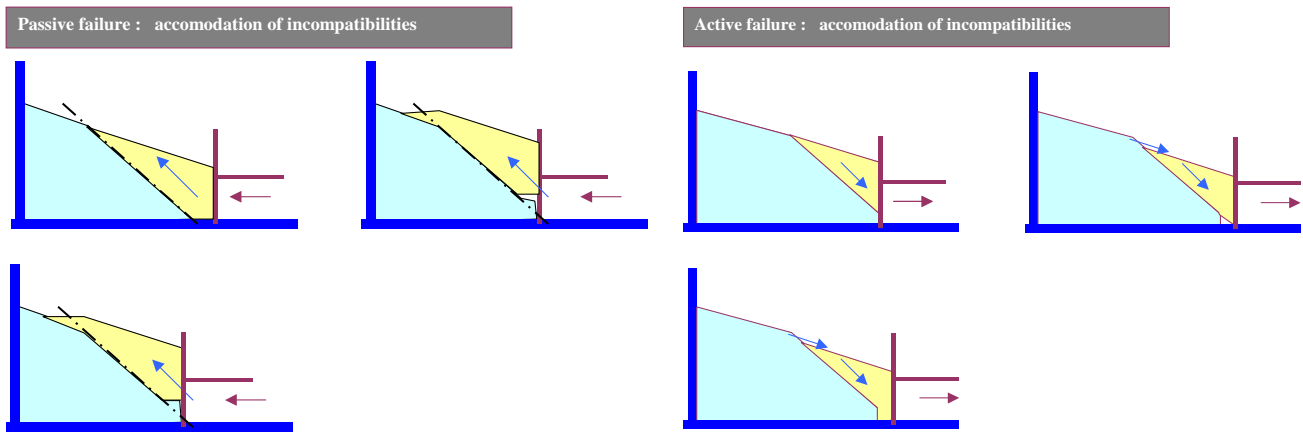

**Figure 8:** Examples of processes allowing to solve the incompatibility of the motion at the bottom right corner. The deformation assumes a constant inclination and position of localisation. The sketch shows that some special mixing shall occur at the bottom corner. This has been observed experimentally, since the direction of flow there is different from elsewhere [3].

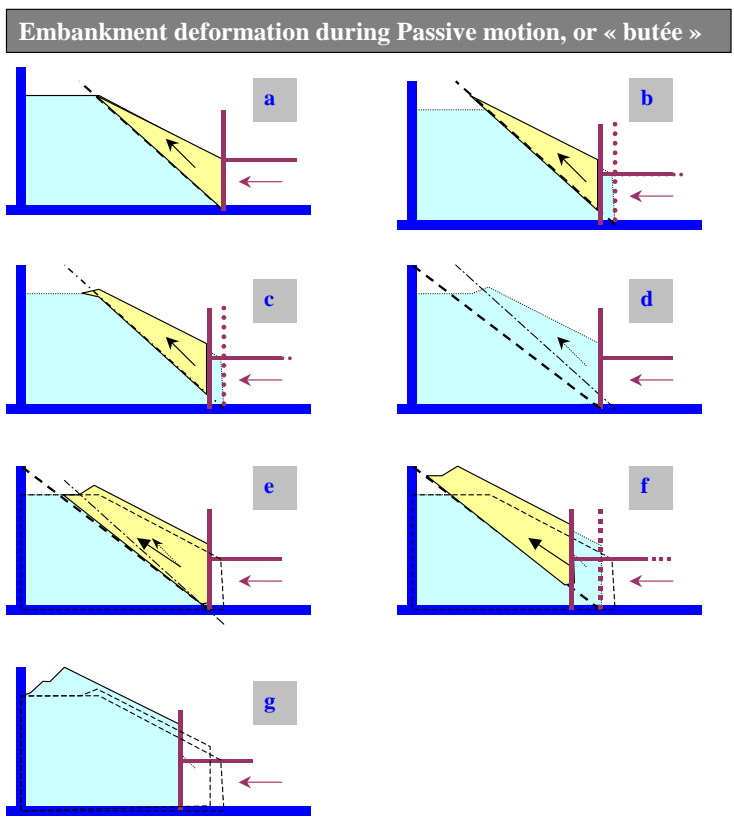

**Figure 9:** Some possible scenario for the upward motion in the passive state :

The scenario is analysed within 7 pictures and considering a bulldozer motion having 2 steps .
  Motion 1: (a, b, c, d)
  Motion 2: (d, e, f, g)

The failure occurs at an angle α which depends on the mean angle of the free surface. As the pile moves up the average slope of the free surface decreases, the angle of sliding decreases, (this is sketched in pictures d & e), and the sliding zone invades deeper the left part of the embankment.

So, as the bulldozer moves, (i) the localisation goes to left and more grains are concerned by the motion, (ii) the direction of motion is less steep.

Figs. 9 and 10 exemplify other parts of the process; they focus on the ways the position and the inclination of the failure zone have to evolve, for both the passive motion (Fig.9) and the active motion (Fig.10). During passive motion, one sees in particular that the bottom left part of the moving zone is set in motion during the only second part of the passive failure. This is due to the fact that the failure starts from the bottom of the bulldozer, so that it enters progressively in the embankment. At the same time the up motion makes the free surface to get less inclined, which forces the failure direction to be less inclined.

During the active motion on the contrary, the free surface gets more and more inclined forcing the failure direction to get steeper; some flow at the free surface may be generated to keep the slope smaller than the friction angle $\varphi$. As the bulldozer goes out during active failure, the failure zone goes out too, because it is locked on the bottom corner of the bulldozer and that this one moves on the right; so the number of grains which are in motion becomes less and less, and their motion is more inclined.

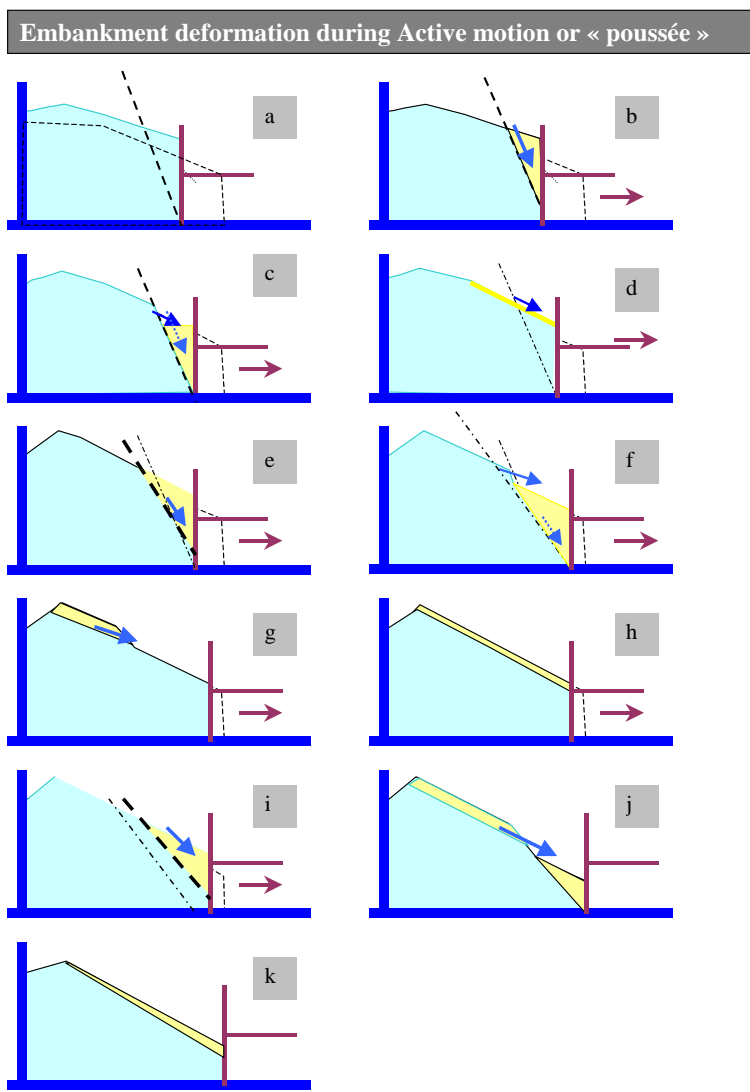

**Figure 10 : Some possible scenario for the downward motion in the active state :**

Downward motion is analysed in 11 images corresponding to 3 backward steps of the bulldozer.
    Step 1: a, b, c, d;
    Step 2: e, f, g, h;
    Step 3: i, j, k;

As soon as the free surface is at the angle of repose, the failure occurs at an angle $\alpha$ which is approximately constant, but the failure zone goes out from the pile so that the sliding zone decreases in size.

As the slope of failure $\alpha$ is larger than the friction angle $\varphi$, a surface flow is generated which maintains the slope at the friction angle (c,d), (g,h) and (j,k). This reshapes the top part of the pile.

However, at the beginning of active motion, the embankment has been compacted by passive motion; so its free surface is less inclined than $\varphi$ at the beginning of the half cycle. This means that $\alpha$ decreases during active motion . this is sketched in pictures (e) and (i) where one see the last angle of sliding together with the present one.

Integrating the motion all along a cycle allows to calculate the average flow:

On what concern the frontiers of the moving zones, one can tell the following: one notes that (i) the grains located in the bulk near the frontier of the mobile zone enters in this mobile zone only at the end of the passive motion, (ii) that they move up left during this motion, and (iii) that they do not move anymore during the active failure. So, the mean motion of the grains in this part is upward left.

Obviously such a permanent flow can only exist if there is a back flow some where else to counterbalance the flux. This forces the convection. Indeed, as shown in Fig. 10, see Figs. 10c, 10d, 10f and 10j, an other permanent flow may be generated at the free surface, which is downwards in this case, and it occurs during the active motion. However, as the directions are not the same, the addition of the two processes can not lead to a complete balance. If the two intensities of flows were the same, the compensation could come from a flow of grains localised at the bulldozer wall and oriented downward. This is indeed observed [3]. However, one sees experimentally a bulk convection too, which means that part of the flow is distributed in volume.

Figure 11 summarises the coupling of the two localised flow during a cycle. However, 2 different cases have to be considered, depending on the value of $\varphi$ and $\varphi_w$, since a periodic steady solution, with the same localisation band in upward and downward motion, exists only within some range parameters $\{\varphi, \varphi_w\}$.

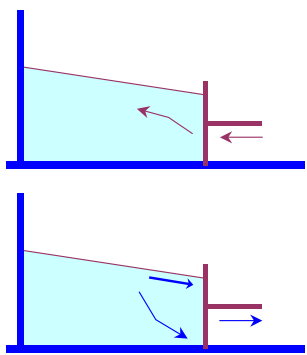

**Figure 11: combined active and passive motions generate convection:**

**During the passive failure:** The sliding direction in the bulk evolves, starting rather inclined and diminishing all along the passive motion. The inner (or more left) part of the pile moves at the end of the passive motion, and does not move back during active failure.

**Active failure** generates a surface flow and the inclination of the free surface keeps increasing till the angle of repose is obtained. Thus the direction of localised bulk sliding decreases.

● ***The case when the passive and active shear bands can get the same inclination***

So when a solution with the same shear band inclination in the passive and the active state can be found the present modelling applies. It assumes a motion by block; as the block shape evolves as a function of the displacement x, this introduces also some boundary effect that generates surface flow at the interface between the moving zone and the non moving one, or at the free surface. But as x is assumed to be small and the condition to be periodic, the inclination of the failure band does not change much during the experiment in the stationary regime; and this inclination is about the same in the active and passive motion; so, one expects that the zone concerned by frontier effects shall occupy a limited part of the block that moves; this is even truer in the limit x→0. Hence one expects that a typical bulk behaviour shall be possible to define in the centre of this block, leading to a homogeneous uniform motion; this is indeed imposed by the hypothesis of localised failure. As the centre of mass of the moving block does not move after a cycle and since the motion is expected to be uniform, the sum of the displacement vector along a cycle shall make a closed loop; this is sketched in Fig. 12: Each point of the central part of the moving block performs the same motion, which is a closed loop. On the contrary, this is not true in the vicinity of the boundaries of the moving block, because these boundaries evolve along a cycle so that the motion becomes intermittent there; in this case the model predicts an average motion: it consists in an up-left mean flow $j_{loc\_band}$ at the bottom of the moving zone, a right-downward flow $j_{free-surf}$ at the free surface; then it shall exist also a rather downward flow $j_{bull}$ near the vertical wall of the bulldozer to ensure no accumulation of grains anywhere in mean. This part of the scheme is sketched in Fig. 12 also.

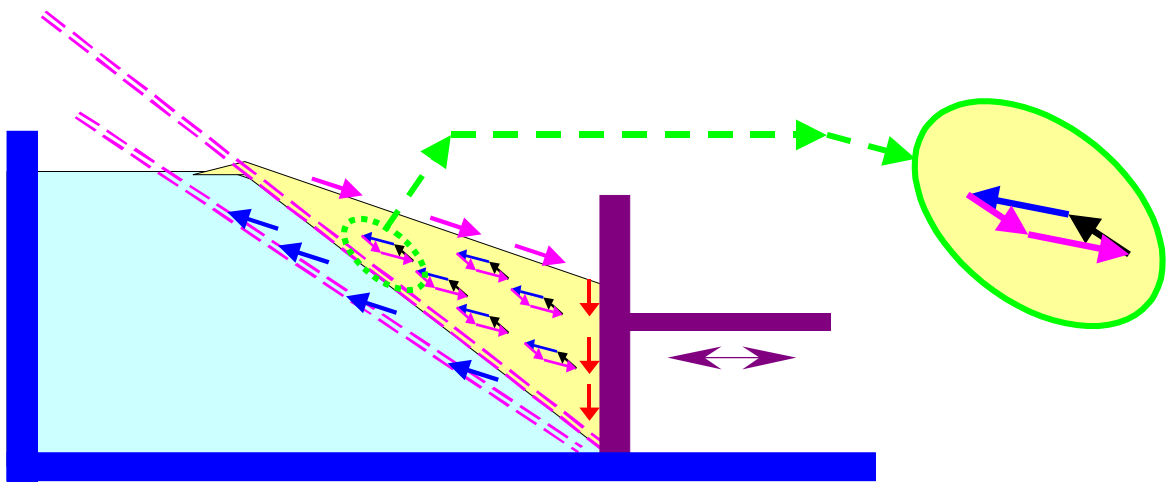

**Figure 12 : Mean flow predicted by the mechanism of localised active + passive failures.**
The modelling assumes that periodic conditions can be satisfied in some range of parameters $\{\varphi,\varphi_w\}$ In the middle of the moving block. Also, the motion is supposed to be homogeneous in the moving block, but to evolve with the phase of the period due to finite amplitude. This leads to the equal closed loops in the centre of the moving block.
This solution is not true in the boundary of the block, since these frontiers evolve with the phase of the period: In the outer left part of the block, a flow exists only at the end of passive motion; this leads to a mean upward flow localised in this region (blue arrows). Also, a mean downward flow is generated at the free surface, during active failure only; this generates a mean downward surface flow (pink arrows). Matter preservation implies that a flow occurs near the bulldozer wall (red arrows).
In the bottom part of the embankment, no flow occurs.

Few questions remain at this stage: if the direction of the flow in the vicinity of boundaries of the moving block has been predicted within this model, the intensities of these flows ($j_{loc\_band}$, $j_{free-surf}$, $j_{bull}$) are not known, but can be calculated in principle from the model. On the other hand, the rule of preservation of matter requires some relationships between these flows, *i.e* $|j_{loc\_band}| = |j_{free-surf}| = |j_{bull}|$. There is no proof at this stage that these relationships are satisfied. If not, it means that a fourth flow shall be also involved to solve the incompatibility, this one can be a bulk flow .

An other point is the following: one sees that the back and forth motion generates a flow localised at the interface of the moving block. But this block is formed of non cohesive granular matter; hence it may deform under perturbation, and this block sees a flow at its surface, which interacts with it. Will not this enforce some spinning of the whole moving block and/or some inhomogeneous flow inside the block?

An other point which remains to be discussed is the real speed of convection, and its dependence upon x. As a matter of fact, the model states that a stationary inclination of the localisation band exists when $x\rightarrow 0$; in this limit of constant inclination the flow shall be 0 at first order. Hence one predicts a flow j that scales as $x^2$ in the limit $x\rightarrow 0$. However, as soon as the amplitude becomes large enough the convection in the moving block is imposed by the displacement at its surrounding, as this displacement changes of orientation during a cycle, and that it stops during part of the cycle, one expects that the mean displacement in the neighbouring zone of the block varies linearly with x at larger x.

• *The case when the passive and active shear bands cannot get the same inclination*

In the range of $\{\varphi, \varphi_w\}$ when one cannot find the same inclination for passive and active yielding, the solution studied above does not hold. Approximately, this occurs when $\varphi>30°$ and $\varphi_w\geq 10°$. In this case the flow becomes more complex and less localised, whose characteristics are different during active and passive forcing. One expects then a convection in the bulk.

## 4. Comparison with experiments:

Fig. 13 reports some data with the large 2d set-up of ref [3.b], ($L_o$=500mm, $H_o$=300mm). The 2d embankment is made of plastic hollow cylinders of outer diameter D=5mm and length L=5mm confined in between 2 vertical glass plates. Most grains have their axis horizontal, but some of them were introduced with their axes parallel to the window in order to break the local hexagonal structure that appears spontaneously with 2d system of equal-size grains. The embankment can be covered by a lid to test the effect of a change of boundary conditions. (The lid was used in the case of experiments reported in Figs 13, 15, 16). The scale of Fig. 13 is in pixels, but it corresponds also approximately to 1 pixel ≈ 1.5mm. The peak-to-peak amplitude $x_o$=40mm of the bulldozer motion is rather large compared to the typical pile height H=230mm; but it is only few times the grain size. This large $x_o$/H ratio explains why the shape of the pile is quite different at the end of passive state (Fig. 13.a.) and at the beginning of passive state (Fig. 13.b).

Indeed, as the free surface shall be a flow line, the upper flow line of Fig. 13 delimits also the free surface (or the lid surface) . So, the free surface flow is rather horizontal in Fig. 13.a, and inclined at the angle of repose or so in Fig. 13.b.

As a matter of fact, the experiment consists in taking automatically video images of the pile at each maximum and/or minimum position of the bulldozer and to track the positions of 3 different grains inserted in the sample; these 3 grains have a colour different from the other ones, which makes them possible to identify automatically with a home-made Visilog program of image analysis. However, one can get some difficulty to identify the complete sequence of positions when the amplitude of motion $x_o$ is large, or where the tracked grains are too numerous; in these cases the trajectories of the grains can exchange erratically, just by interchange of labels, when their position become too nearer from each other; this perturbs the analysis, by introducing large jumps in the trajectory; this effect is quite difficult to avoid and is amplified also by some residual non homogeneity of the light, by interfering reflection of light, causing a non uniform distribution of the grey scale in the pixels.

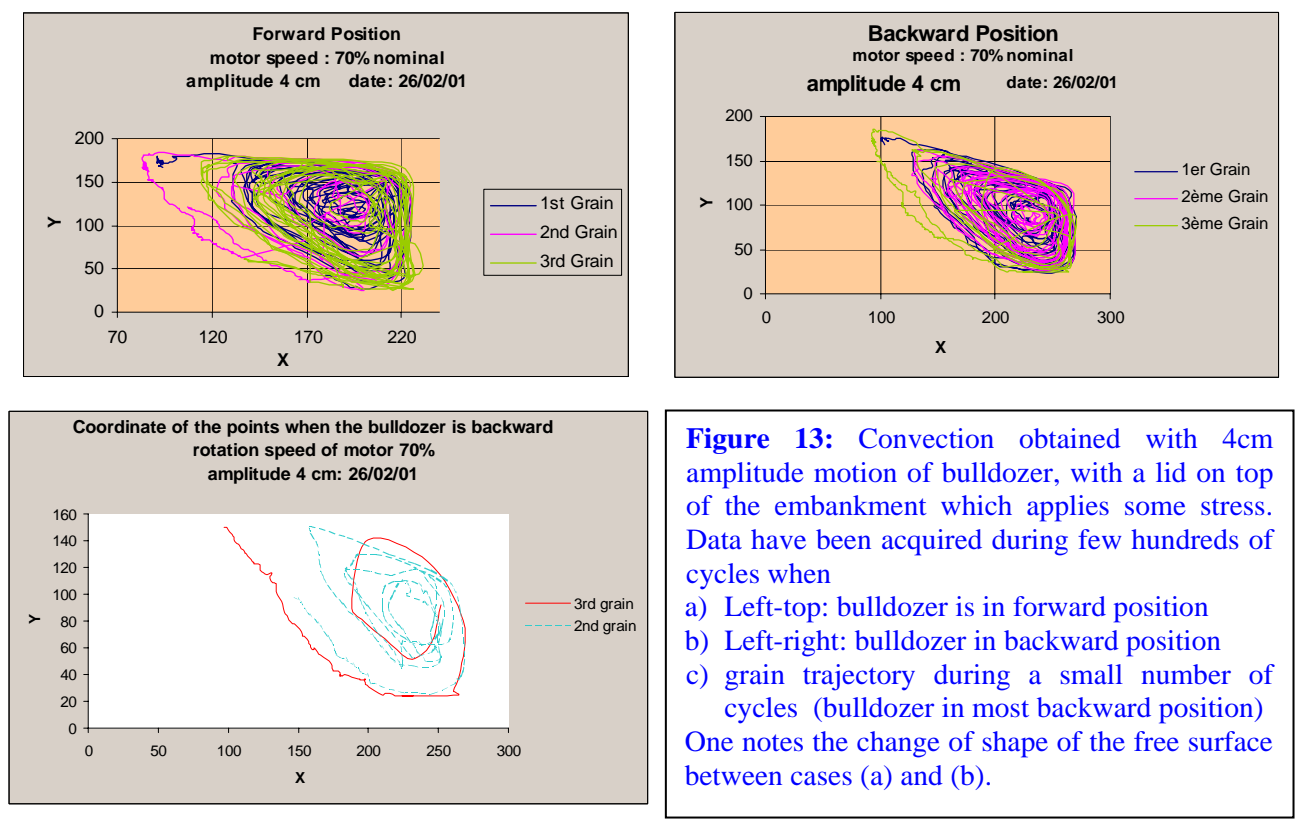

**Figure 13:** Convection obtained with 4cm amplitude motion of bulldozer, with a lid on top of the embankment which applies some stress. Data have been acquired during few hundreds of cycles when
a) Left-top: bulldozer is in forward position
b) Left-right: bulldozer in backward position
c) grain trajectory during a small number of cycles (bulldozer in most backward position)
One notes the change of shape of the free surface between cases (a) and (b).

An other problem is the optical aberration generated by the optics of the camera; it deforms the field of view, which generates a systematic error on the real position recorded. However the defect can/could be automatically corrected by the computer; it requires simply to calibrate the experiment by recording first the image of a regular lattice of regular points. Fig. 14 gives this calibration; it is the disturbed image recorded by the program of a regular square lattice of points. The results reported here after are not yet corrected from optical aberration, while they can be using Fig. 14 data if needed.

As told already, Fig. 13 reports the successive positions of the 3 different grains as a function of the cycles during more than 370 successive cycles. One can see some examples of exchange of grain trajectories, on the green trajectory in Fig. 13b for instance: they correspond to sudden horizontal jump.

Besides these sudden jumps which are artefacts, one observes also a discrete motion made of quite small increments between the successive positions of the grains. This series of point looks like a continuous trajectory, the discreteness being due to the discrete time set. So the trajectory looks rather smooth when determined at a constant phase of the cycle. However, it is not completely smooth and regular as shown in Fig. 13.c, which reports 2 short parts of the trajectories extracted from Fig. 13.b. So, one sees also that the grain motion is slightly erratic, as if it was submitted to a bias random walk. This erratic motion is important, since it allows the exploration of the bulk by the grain in the triangular region where convection occurs: For instance one can see that any of the 3 grains of Fig. 13.b has explored the same volume at the end, when one waits long enough. In conclusion the convection flow is perturbed by diffusion. But the volume where the grains diffuse seems restricted to the top right part, where the convection is confined.

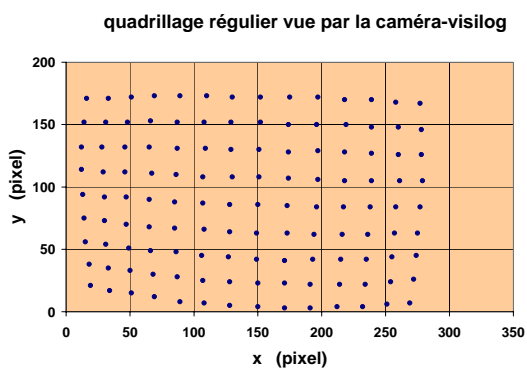

**Figure 14 : Optical image recorded by the camera of a regular array of points dispersed in the embankment;** image has been taken with the same magnification and same aperture as other frames. The deformation is due to spherical aberrations due to the optics.
Inverse transformation of this picture allows to reconstruct the correct observation of all observations.

So a question arises: are convection and diffusion strictly coupled? In other word, how convection flow and diffusion speed scale with amplitude $x_o$ of motion, are they proportional always ? This is not yet analysed due to the lack of data. And the problem is complicated:

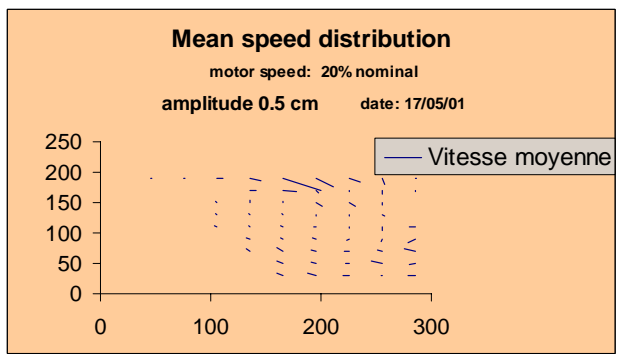
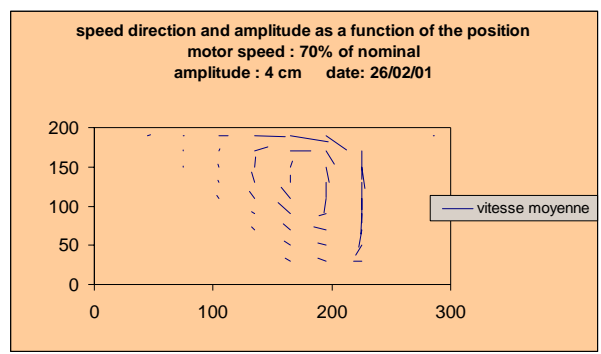

**Figure 15 :** Distribution of speed of the mean flow in an embankment submitted to cyclic bulldozing of different amplitude $x_o$: $x_o$= 0.5cm (left) and $x_o$= 4cm (right). These data come from the analysis of flows similar to the ones of Fig. 13.
One notes that the two distributions are spatially different.

Fig. 15 reports the variations of the convection speed for 2 different amplitudes $x_o$, *i.e.* $x_o$=5mm in Fig. 15.a, and $x_o$=40mm in Fig. 15.b. This figure has been determined by (i) following each grain independently along time, (ii) by determining the difference of positions of the same grain after few cycles, at the same phase of the cycle, (iii) by recording this vector displacement as a function of the position of the grain at the time of the measurement and then (iv) averaging this vector over the different examples encountered all along time in a small region of the embankment, i.e. when the grain passes back within the vicinity of the tested zone. According to this Figure, one sees that the distribution of mean speed is non homogeneous, and that the shape of the spatial shape of the distribution evolves with $x_o$, so that the mean amplitude of the convection evolves too…. This makes the system rather difficult to analyse, and the trends difficult to define.

One sees also in Fig. 15.b that the average speed is much faster near the free surface and near the wall than near the localisation. This is an effect of confinement of the flow at this location: because the convection occurs in the bulk, and due to the preservation rule of matter, the more extended the section S of the flow the smaller the speed v since $<j>=<Sv>$ shall be constant along the flow lines.

On the contrary, this trend is not displayed in Fig. 15.a. As the preservation rule shall be satisfied whatever the experimental conditions are, it means that the experimental uncertainty of Fig. 15.a is quite large for this small amplitude $x_o$=5mm of vibration . Indeed, $x_o$=5mm corresponds to the size D of a grain.

**Figure 16 :** Motion during passive and active failures, followed during few tens of cycles.
(a) and (b): 4cm amplitude,
(c) : 1cm amplitude.

(a): top-right fig., (b) bottom-right, (c) bottom-left

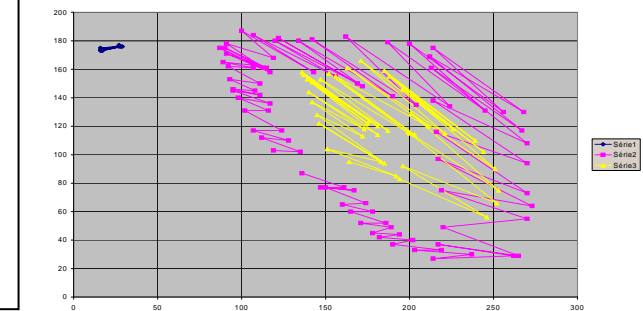
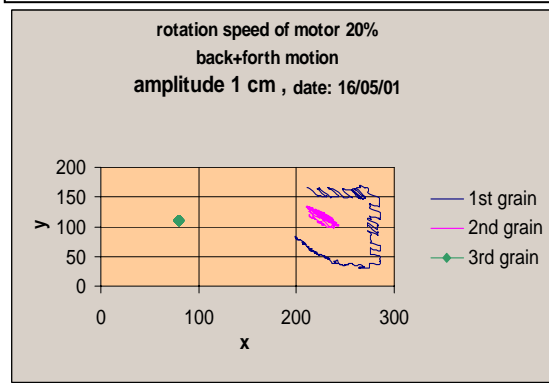
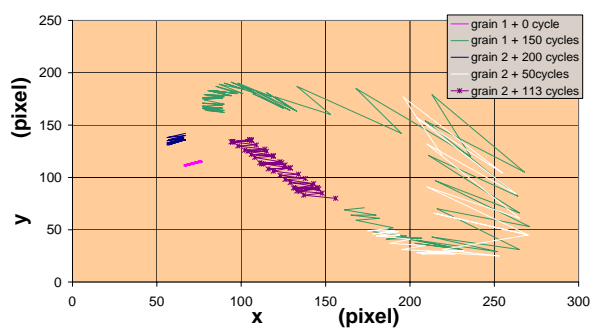

The level of uncertainty may explain partly why the geometry of flows is different at different amplitude : so, correct comparison requires to improve much the experiment accuracy first. In the present condition the difference observed have to be confirmed.

In the same way, comparing the convection speeds and the diffusion speed and their variations with the amplitude $x_o$ is not accessible yet, since the experiment uncertainty is too large. So the linear (or non linear) scaling of diffusion time with convection speed cannot be identified at this stage. It requires at least much more data.

However answering this question is a major goal, since it may teach about the connection on how local and global motions are connected together in granular matter, how the randomness generated at local scale by the granular structure propagates to larger scales and to the whole-system size; and how the propagation depends on the intermediate scale $x_o$.

Fig. 16 reports the successive positions of few grains taken both at the maximal and minimal displacements of the bull-dozer, for 2 different amplitudes of motion $x_o$, *i.e.* $x_o$=1cm and $x_o$=4cm. This experiment allows to determine the mean sliding direction of the grains during the back- and the forth- motions of the bulldozer. In the Coulomb approximation, this shall correspond to the active and passive directions of sliding. In the Coulomb approximation too, the active and passive sliding is localised; hence the grain motion shall be uniformly distributed in some region of the bulk. Indeed, Fig. 16 demonstrates the contrary because the directions of sliding is non homogeneous: First in the vicinity of the localisation band the amplitude of motion is much shorter than near the bulldozer wall. Also the direction of sliding varies with the position in the embankment; for instance, in the vicinity of localisation the motion due to active failure is horizontal; this is not at all predicted by the modelling. Also the direction of passive failure is always steeper than the direction of active failure; this is true except in the vicinity of the bulldozer wall where it is the contrary.

In Fig. 16.b, the trajectory of the grain has been cut into different pieces, to avoid artefact due to intermixing of trajectories. Comparing Fig. 16.a and 16.b, one observes that horizontal amplitude of motion is approximately the same for the pink trajectory of Fig. 16.a and brown trajectory of Fig. 16.b, while the vertical motion seems larger. This may surprise; however the position in the bulk are slightly different, the "brown Fig-16.b grain" being located deeper in the pile than the "pink Fig-16.a grain". This is an effect due to shear localisation and to the localisation of flow: one expects that the flow is non homogeneous in the vicinity of the boundaries where convection is restricted,.

**Pseudo elastic behaviour:** One sees also on the left part of Fig. 16.a a grain which moves back and forth periodically staying at the same location in mean. This shows a behaviour which looks elastic, and which is quite different from what occurs in the other part of the material. Similarly, the blue and pink trajectories of Fig. 16.b are nearly periodic. However, looking carefully at the experiment during it works tells that true elasticity is not involved in the mechanism, since the grains move by block in this region and that sliding occurs along a localised plane localised further on the left of the grain; this localisation surface and motion is exactly the same for back and forth motion except quite rarely and intermittently, so that the system evolves very slowly under repeated bulldozer motion. Indeed, in this zone the motion amplitude is few grains at most so that the organisation of the sliding surface is only perturbed by intermittence in this direction. This leads to the observed pseudo-periodic motion, which looks rather elastic, but which is not, since the deformation itself is not elastic because some of the grains slide and roll at some location. We call it pseudo elastic behaviour.

So, this allows to conclude that quasi-periodic behaviour is caused by a pseudo elastic compression of the assembly during passive motion. As the compression is mainly horizontal, the grain motion due to pseudo elastic deformation is also mainly horizontal. This pseudo elasticity may explain also the observation made in the last but one paragraph, *i.e.* the relative independence of the horizontal component of motion for the "brown Fig-16.b grain" and the "pink Fig-16.a grain", while their vertical components are rather different : in this zone of the embankment one expects the fully plastic response to be very sensitive to the real position, while the pseudo elastic one shall not, because it is more extended spatially; hence, as the pseudo elastic response is large in this zone compared to the plastic one and as the plastic response evolves fast this explains the observed fact.

This means also that the experimental data cannot be directly compared to the modelling of section 2, since no pseudo elastic response was considered there. One has first to subtract the elastic response from the experimental data.

One expects also that performing such a correction in the zone located in the region of localised yielding leads to a step-like motion parallel to the direction of localisation. This is indeed true within the experimental error bar.

At last the pseudo elastic behaviour which has been described in the previous paragraphs seem to be just an example of what we have tried to model: When amplitude of motion is small enough and when periodic forcing is used one shall expects to get a periodic plastic process. Hence, the pseudo elastic behaviour encountered here is just the manifestation of this periodic plastic process. This effect which seems then to be in contradiction with a Coulomb-like approach at first sight, is in agreement with our own modelling in fact.

A question remains: how does the pseudo elastic behaviour depend on the regularity of the packing structure? In other words, does it depend only on $x_o/H$ alone, or does $x_o/D$ intervene too?

An other point which is worth noting is the fact that using a lid of few grams on top of the pile modifies importantly the distribution of stress since the orientations of the failure zone and of the "free" surface become both less steep. This is expected from calculation. It is worth recalling that the present data (Figs 13,15,16 correspond to experiments with a lid.

## 5. Discussion and Conclusion

It has been shown in this paper that convection flow can be generated in a granular medium submitted to quasi static periodic bulldozing. The Coulomb approach which has been used, which assumes localised deformation, is able to explain, qualitatively at least, the phenomenon. In particular, the inclinations of the localised-strain yield and of the free surface has been found just by imposing steady periodic solution; the model gets also the right direction of the main convection and it predicts some localised flow in the bottom part of the pile near the bulldozer, which is just due to an incompatibility of the deformation process at this level.

However we do not know if the present modelling explains the intensity of the flow. Also, the present modelling does not catch completely the inhomogeneous nature of the flow in the bulk because it assumes a motion by blocks. Anyhow, existing experimental data are too imprecise to allow a quantitative test of the modelling. Also, the fact that the flow occurs in the bulk instead of being localised can be due to the range of the experimental friction φ and $φ_w$ of experiments, because it has been demonstrated in Fig. 5 and Table 1 that no periodic steady localised solution can be found in some range of parameters {φ, $φ_w$}, and since this range corresponds very often to experiment, *i.e.* {φ>30°, $φ_w$>10°}. In this case, the flow shall be not localised, and more complex and more inhomogeneous.

- *Small strain but large stress change*

It is worth noting that while the system studied here imposes small deformation, it requires very large variation of the applied forces and stresses; this is obvious from Table 1, which shows that the ratio between active and passive forces P can be as large as 100 in some range of friction parameter {φ, $φ_w$}. So one cannot consider the system as simply "submitted to small cyclic perturbation". This is true for strain, but not for stress.

As recalled in the introduction, the case of small deformation is considered as the domain of stress-strain rheology of foundations; hence it is not the case "*stricto sensu*" here. However, the present example shows that imposing large periodic change of stress may generate important macroscopic flow after a while. An other domain where similar effects could be found is the one of the perturbation of foundations during seisms, siçnce seisms generate important periodic variations of stress.

- *Parallel with acoustic streaming in fluid mechanics*

It is also worth making the parallel between this flow generation and the one met when a fluid is contained in a container which is subjected to sonic or ultrasonic waves [9-11]. In this case, the oscillation of the wall forces the liquid to flow also. The flow obeys inertia in the middle of the sample, but it is attached to the wall due to viscous effect, so that it is not homogeneous. Let us now assume the wall is not perfectly flat, the flow line shall adapt to the geometric constrain. This generate gradient of speed and space dependent dephasing. This is specially true near a corner; but it is also encountered for a sinusoid bottom,…. So, be **u**(x,t) the flow of the fluid with time t, one has ∫u(t)dt=0 over a period; be also a particle of fluid its speed **w**(x,t) at time t is without approximation

$$\mathbf{w}(x_o,t)=\mathbf{u}(x_o+\int_{t_o}^{t}\mathbf{w}(t)dt,t) \qquad (5)$$

leading to

$$\mathbf{w}(x_o,t)\approx\mathbf{u}(x_o,t)+\{[\int_{t_o}^{t}\mathbf{w}dt]\nabla\mathbf{u}(x,t)\}_{x=x_o} \qquad (6)$$

which has a non zero mean, due to the variation of phase with space. In other words, because the particle move, its mean motion is non zero, due to the variation of phase with space. This forces a coherent stream, which takes place in the zone of liquid "attached to the boundary". This zone is called the boundary layer.

In the same way, Fig. 16 shows that the grains perform a systematic drift after each period; this drift depends on the location, due to the boundary conditions and their variations. This generates the stream. Hence, there is a strong parallel between the two mechanisms.

However, there are at least two main differences remain between them, which are both linked to the difference of nature between viscous and solid frictions: in the case of acoustic streaming in liquid, the viscosity ν ensures a non sliding condition at the boundaries, which is not the case with grains. So it ensures firstly that the flow speed is zero at the vibrating wall, which is not the case in Figs. 13, 15 and 16. Secondly, the convective flow is generated due to Eq. 6 during acoustic streaming near the vibrated wall, in the boundary layer thickness , which has a small extension $\delta \approx (\nu/\omega)^{1/2}$ at high frequency [9-11]. But this primarily flow generates in turn by viscous friction a secondary flow in the fluid bulk itself, which rotates in contrary direction. In the case of solid friction, the primary flow extends to a much larger zone scaling as H*H (instead of H*δ), and the secondary flow does not exist, since the left part of the embankment is quiet.

- *Statics versus Dynamics excitation*

It is also worth noting that the flow pattern of Fig. 13 is observed similarly in granular media in container under "dynamic" horizontal excitation, *i.e.* intense horizontal vibration of large frequency [12] (f>15Hz). In this case, the mode of excitation is not quasi-static any more (for instance one observes a periodic separation of the medium from the vertical wall of the container). Anyhow, it seems that the stream results from the same combinations of mechanisms as those which provoke the quasi-static streaming of Fig. 13. Some difference shall probably remain between the two modes, *i.e.* quasi-static and dynamic, because the two stackings are submitted to different boundary conditions all along the cycles. Nevertheless, one can guess that some rather good description shall be obtained using local quasi static rheology (stress-strain relation) with correct boundary conditions.

- *Role of Fluctuations*

Indeed, the experiment of Fig. 13 generates large fluctuations because the grain size is large compared to the sample size. Does it mean that fluctuations play an important part in the streaming mechanism? It is not obvious for the following reason: First, Fig. 13 shows that flow rules do not fluctuate so much. Secondly, we can try and compare this case with what occurs in the case of acoustic streaming in viscous liquid: in this case a similar effect will occur ; and it will be even much enhanced, because atoms of liquid move randomly with a speed much faster than the one of the convection flow, as far as the flow remains much smaller than the sound speed. This is why we believe the stream of Fig. 13 is controlled completely by the macroscopic mechanics, and by the plastic behaviour of the granular material. Local fluctuations generate diffusion, which is a much less efficient process of motion than convection at large scale, but which is essential for 2d mixing.

- *Other parallel with the mechanics of fluids: chaotic advection*

One can push further the parallel with the problem of steady flow in an unspecified fluid [13-15]. Let us look at the motion over several periods; in mean it is a 2d steady flow at constant volume; this imposes div(**v**)=0, where **v** is the mean local velocity of a particle during a cycle; this imposes [10] that it exists a stream function Ψ such as:

$$v_x = dx/dt = \partial \psi / \partial y \tag{7.a}$$

&

$$v_y = dy/dt = -\partial \psi / \partial x \tag{7.b}$$

where ψ is called the stream function.

Formally, the mathematical problem becomes equivalent to the problem of a particle in a field, which is managed by the system of equation:

$$dp/dt = \partial H / \partial q \tag{8.a}$$

$$dq/dt = -\partial H / \partial p \tag{8.b}$$

where H is the Hamiltonian of the particle. It is known that this problem is integrable and that it cannot lead to chaos in 2d. Thus, a permanent two-dimensional flow of an incompressible fluid cannot give chaos and cannot be used for active mixing. Within other words, the permanent flow lines which results from such a device can be plotted in a 2d plane; but these flow lines cannot cross, because steady stream lines cannot cut one another, otherwise the same point could flow in two different manner; so due to the 2d topology, these lines shall look either as concentric or as open (if the system is open); they may separate into few distinct zones, *i.e.* vortices; so the typical distance between two concentric flow line scales then as the vortex size, which is also the device size L itself, most often. So, when used for mixing such 2d device can only mix with the help of Brownian motion; this process takes place at the grain scale D and its time efficiency vanishes as $(D/L)^2$ when the vortex expands over the whole cell, or as $(D/L_{vortex})^2$ when the flow separates into few vortices.

This is just to remind that the experimental set-up of Fig. 12 cannot be used as a mixer, because its efficiency decreases strongly when the grain size is decreased and/or the bulldozer height is increased. Indeed, the topology argument, on the structure of the flow line, does not remain with 3d devices generating fully developed 3d complex flow lines, or in 2d with intermittent flow. So, good mixing requires fully developed 3d flow pattern, which requires at least 2 directions of excitation with 2 different frequencies.

It is worth reminding that rotating cylinders pertain also to the category of 2d device; so they can never be good mixers. So, it is quite surprising that they are used in plants so often. It is just as if one wanted to dissolve a sugar in a cap of coffee just by rotating the cap, without using a spoon; this is just inefficient.

On the contrary, classic concrete mixer are classic examples of 3d flow generators; for this reason, they are much better mixers.

An important domain of application for granular matter is then mixing and segregation [16].

- *Conclusion*

As a conclusion of this study, it is worth drawing few consequences. For convenience, I will term here after "soil mechanics" the classic stress-strain laws a granular material obey in the quasi static regime. Of course, if "soil mechanics" exists, it shall apply to a much larger domain than the one of soils.

→ *$1^{st}$ Interest of the experiment: test of validity of the "mechanics of soil" to describe flowing:* Indeed it has been demonstrated that quasi static cyclic deformation can generate flows in granular samples; we have now to prove that the deformation process obeys classical law of plastic deformation of soil. This is done in Fig. 17, which reports the experiment of Fig. 12 made in the spirit of "soil mechanics": the packing is made of two kinds of duralumin cylinders ($d_1$=3mm, $d_2$=5mm, l=50mm), and the deformation field is visualised using a superimposed square lattice. There is no lid in the present case, and the experiment starts with the steady configuration. Indeed, one can observe the wedge motion, the failure zone, as in classic text book; but one observes also the rotation of the lattice in the middle of the wedge, and its deformation, which are the consequences of the flow pattern of Fig. 13.

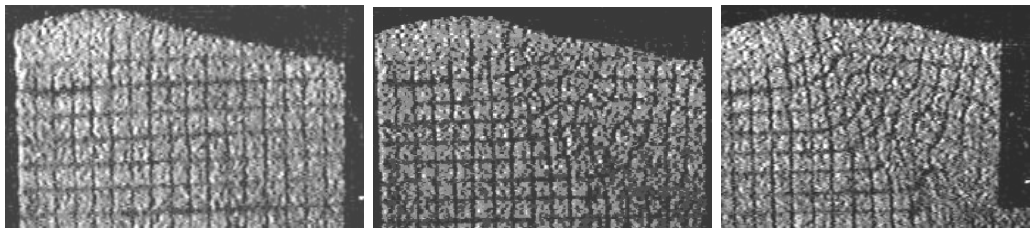

**Figure 17:** Bulldozing experiment similar to the one of Figs. 12 & 13, obtained with a larger embankment and heavier grains of smaller sizes; the initial conditions correspond to a pile to which cyclic forcing has been applied already for a while. (left) initial conditions, i.e. not deformed pile; (middle) after 1 cycle; (right) after 1.5 cycles. One observes the localisation of the deformation at the base of the wedge; the squares in the centre of the moving wedge has rotated in the right photograph, and the lattice has deformed, demonstrating the existence of a complex non homogeneous flow.

→ *$2^{nd}$ Interest: Validity of the statistical approach in the mechanics of piles:* This experiment shows that cyclic solicitation forces the granular medium to flow; hence it forces the grain configuration, the contact distribution and the local-force network to change. Let us now consider a homogeneous sample and apply to it some cyclic deformation; as it is cyclic, the shape of the sample remains the same; but the flow which is generated allows grain mixing; hence it allows to apply some statistical approach; in particular, due to this mixing the principle of the most probable state shall be valid likely. On the other hand, if this principle applies to the above situation, why shall not it be valid at the earlier stage, or with an other pile? Since no demon has built the pile *a priori*.

→ *"Fragile" behaviour and the role of force- and contact- networks* [17, 18]. Indeed one can still refuse applying the principle of the most probable state and still argue that the contact and force network are quite inhomogeneous, which reveals the heterogeneous nature of the mechanics. Is this a contradiction really? One notes that the heterogeneity exists at the microscale; but at larger scale, one speaks already in terms of deformation and stress, *i.e.* in terms of averages which have macroscopic meaning. Furthermore if the packing was completely regular one may tell that statistical approach would not apply to it; but its behaviour would not be regular. Of course, a disordered material would behave as a glass if the system of force was frozen, but as it deforms easily, its underlying networks of contacts and forces evolve; so if their evolutions are fast enough, they validate the description in terms of means; this may justify the existence of a stress-strain relation and the efficiency of statistical description. In this case, local force should be related to the mean stress via a principle of maximum entropy; this seems to be satisfied [19].

On the contrary, if all the forces were equal, the packing would not be able to adapt itself to some small change of stress and would break spontaneously under some specific stress solicitation; this would lead to some "fragile" behaviour [18] .

Turning now to smaller samples, one shall expect that the number of completion becomes smaller, so that the system becomes unable to adapt itself to some stress change; this predicts some noise on the stress-strain curves, and the smaller the sample, the larger the noise. This is exactly what can be observe with triaxial test experiment [20] or in numerical simulations.

One shall note on the contrary that some stress-strain curve exhibit spontaneous stick-slip. This phenomenon can be seen as revealing the "fragile" nature of the packing; it occurs in some samples, probably induced by cohesion. In the case we have studied [20], we have shown that this fragile nature was enhanced by the macroscopic behaviour which becomes periodic for samples containing $10^9$ grains or more.

→ *Use of quasi static law of deformation:* So, "soil mechanics" laws shall be able to predict the complete flow of Fig. 13, within many details; but it is probably quite hard. We see for instance in Figs. 13 & 16 that the flow is primarily generated by large localised deformations; this is rather simple to compute. But these deformations provoke the flow in the bulk; in particular it provokes the slow rotation of the centre of the triangle while the triangle preserves its orientation and shape; these are drastic constrains. So it is a real "technical challenge" for soil mechanics programs. Also periodic conditions have to be found for the stress near the shear band and in the triangle …. The use of series expansion with two time scales may help probably; the rapid one would correspond to the cycle and would be used to describe the localisation, while the slow time would help describing the flow, as it is commonly used in fluid mechanics; but the main problem remains, that is to identify the evolution of the boundary conditions, to identify the true stress distribution and its periodic evolution. This is quite a technical challenge.

Indeed, this experiment helps demonstrating that the limit between solid-like and liquid-like behaviours is not simple; it is mainly a question of points of view. Furthermore, the difference comes from the macroscopic treatment, hence from macroscopic equations; so it has nothing to do with microscopy; this explains why discussing about Brownian motion of the particles does not help.

At last, this experiment helps demonstrating that the limit between static and dynamic is not discontinuous, because vibrating faster than 5-10Hz will not modify efficiently the characteristics of the flow. Also this experiment may have some application in the domain of security of foundations against earthquakes and of foundations against liquefaction of soil.

*Acknowledgements :* CNES is thanked for partial funding.

*Notations :*

The following notations are used: $tg(\xi)=\sin(\xi)/\cos(\xi)$ is the tangent of $\xi$; $cotg(\xi)=1/tg(\xi)$; $x$ is the horizontal displacement of the bulldozer; $x_o$ refers to the amplitude of back and forth motion.

The electronic arXiv.org version of this paper has been settled during a stay at the Kavli Institute of Theoretical Physics of the University of California at Santa Barbara (KITP-UCSB), in june 2005, supported in part by the National Science Fundation under Grant n° PHY99-07949.

*Poudres & Grains* can be found at :
http://www.mssmat.ecp.fr/rubrique.php3?id_rubrique=402